\documentclass[11pt,showpacs,showkeys,preprintnumbers,nofootinbib]{revtex4-1}

\usepackage{amssymb,amsmath,graphics,graphicx}

\begin{document}

\newcommand{\pderiv}[2]{\frac{\partial #1}{\partial #2}}
\newcommand{\deriv}[2]{\frac{d #1}{d #2}}

\title{Non-Equilibrium Phase Transitions Induced by Social Temperature in Kinetic Exchange Opinion Models on Regular Lattices}

\author{Nuno Crokidakis}
\thanks{nuno@if.uff.br}

\affiliation{
Instituto de F\'{\i}sica, \hspace{1mm} Universidade Federal Fluminense \\
Av. Litor\^anea s/n, \hspace{1mm} 24210-340 \hspace{1mm} Niter\'oi - Rio de Janeiro, \hspace{1mm} Brazil}

\date{\today}

\begin{abstract}
\noindent
In this work we study the critical behavior of a three-state opinion model in the presence of noise. This noise represents the independent behavior, that plays the role of social temperature. Each agent on a regular D-dimensional lattice has a probability $q$ to act as independent, i.e., he can choose his opinion independent of the opinions of his neighbors. Furthermore, with the complementary probability $1-q$ the agent interacts with a randomly chosen nearest neighbor through a kinetic exchange. Our numerical results suggest that the model undergoes nonequilibrium phase transitions at critical points $q_{c}$ that depend on the lattice dimension. These transitions are of order-disorder type, presenting the same critical exponents of the Ising model. The results also suggest that the upper critical dimension of the model is $D_{c}=4$, as for the Ising model. From the social point of view, with increasing number of social connections, it is easier to observe a majority opinion in the population.

\end{abstract}


\maketitle

\section{Introduction}

In the last years, several models of opinion dynamics were studied in order to analyze social phenomena like polarization, extremism, conformity, and others \cite{galam_book,sen_book}. Indeed, social systems are interesting even from the theoretical point of view: they exhibit rich emergent phenomena, that results from the interaction of a large number of agents. This interdisciplinary topic is usually treated by means of computer simulations of agent-based models, which allow us to understand the emergence of collective phenomena in those systems.

The impact of conformity/nonconformity in opinion dynamics has attracted recent attention of the physicists \cite{galam,lalama,sznajd_indep1,sznajd_indep2,sznajd_indep3,javarone1,javarone2,meu_indep,meu_pmco}. Nonconformity behaviors like anticonformism or independence, that are introduced in opinion models as disorder or noise, lead to the occurrence of phase transitions, making the models more realistic since distinct opinions can coexist in the population, as occurs usually in referendums and elections \cite{galam_book,sen_book}. The independent behavior was discussed recently in opinion models \cite{sznajd_indep1,sznajd_indep2,sznajd_indep3,meu_indep,meu_pmco}. Independent individuals tend to resist to the groups' influence, taking your own opinions regarding a subject independent of the other individuals. Independence is a kind of nonconformity, and it acts on a opinion model as a kind of stochastic driving that can lead the model to undergoes a phase transition. In fact, independence plays the role of a random noise similar to social temperature \cite{lalama,sznajd_indep1,sznajd_indep2,sznajd_indep3,meu_indep}. 

In this work we study the impact of independence on agents' behavior in a kinetic exchange opinion model defined on regular D-dimensional lattices. For this purpose, we introduce a probability $q$ of agents to make independent decisions. In the absence of noise the population reaches consensus with all agents sharing one of the extreme positions. Our numerical results suggest that the model undergoes phase transitions at critical points $q_{c}$ that depend on the lattice dimension. These transitions are of order-disorder type, presenting the same critical exponents of the Ising model within error bars. The results also suggest that the upper critical dimension of the model is $D_{c}=4$, as for the Ising model.

The organization of the work is as follows. In Section 2 we present the microscopic rules that define the model. In Section 3 the numerical results are discussed for the model defined on dimensions $D=2, 3$ and $4$. Finally, our conclusions are presented in Section 4.


\section{Model}

Our model is based on kinetic exchange opinion models (KEOM) \cite{lccc,p_sen,biswas11,biswas}. A population of $N$ agents is defined on a regular D-dimensional lattice of linear size $L$ with periodic boundary conditions, i.e., we have $N=L^{D}$. Each agent on a given lattice site $i$ carries one of three possible opinions (or states), namely $o_{i}=+1$, $-1$ or $0$, and he/she can interact only with his/her nearest neighbors. The following rules govern the dynamics:

\begin{enumerate}

\item A lattice site $i$ is randomly chosen;

\item With probability $q$, the agent on site $i$ will act independently. In this case, with probability $g$ he/she chooses the opinion $o_{i}=0$, with probability $(1-g)/2$ he/she adopts the opinion $o_{i}=+1$ and with probability $(1-g)/2$ he/she chooses the opinion $o_{i}=-1$;

\item With probability $1-q$ we choose at random one of the $z=2D$ nearest neighbors of site $i$, say $j$, in a way that $j$ will influence $i$. Thus, the opinion of the agent $i$ in the next time step $t+1$ will be updated according to
\begin{equation}\label{eq1}
o_{i}(t+1) = {\rm sgn}\left[o_{i}(t) + o_{j}(t)  \right]\,,
\end{equation}
where the sign function is defined such that ${\rm sgn}(0)=0$.
\end{enumerate}

In the case where the agent $i$ does not act independently, he/she can change his/her state following a rule similar to the one proposed recently in a mean-field KEOM \cite{biswas}. Notice, however, that in Ref. \cite{biswas} the two randomly chosen agents $i$ and $j$ interact with competitive couplings, i.e., the kinetic equation of interaction is $o_{i}(t+1) = {\rm sgn}\left[o_{i}(t) + \mu_{ij}\,o_{j}(t)  \right]$. In this case, the couplings $\mu_{ij}$ are random variables presenting the value $-1$ ($+1$) with probability $p$ ($1-p$). In other words, the parameter $p$ denotes the fraction of negative interactions. In the mean-field case, the model with competitive interactions \cite{biswas} presents an order-disorder transition at $p_{c}=1/4$, with the same exponents of the mean-field Ising model, namely $\beta=0.5$, $\gamma=1$ and $\nu=2$ \footnote{The discrepancy on the exponent $\nu$ was observed in other KEOM \cite{meu_indep,meu_pmco}, and was associated with a superior critical dimension $D_{c}=4$, that leads to an effective exponent $\nu^{'}=1/2$, obtained from $\nu=D_{c}\,\nu^{'}=2$}. In the case of a square and cubic lattices, the KEOM with competitive interactions was studied recently: it undergoes a nonequilibrium phase transition at $p_{c}\approx 0.134$ (for the two-dimensional square lattice) and $p_{c}\approx 0.199$ (for the three-dimensional cubic lattice), and in the absence of negative interactions ($p=0$), the population reaches consensus states with all opinions $+1$ or $-1$ \cite{mukherjee}. For $p_{c}\leq p\leq 1.0$ the society is in a paramagnetic disordered state, with an equal fraction of the two extreme opinions $+1$ and $-1$ (on average).

Thus, our Eq. (\ref{eq1}) represents the D-dimensional version of the KEOM of Ref. \cite{biswas} with no negative interactions, i.e., an extension of the model presented at mean-field level in \cite{meu_indep}. The above parameter $g$ can be related to the agents' flexibility \cite{sznajd_indep1,meu_indep,meu_pmco}. In this case, for $q=0$ (no independence) all stationary states will give us $m=1$, where $m$ is the order parameter of the system, 
\begin{equation} \label{eq2}
m = \left\langle \frac{1}{L^{D}}\left|\sum_{i=1}^{N}\,o_{i}\right|\right\rangle ~,
\end{equation}
and $\langle\, ...\, \rangle$ denotes a disorder or configurational average taken at steady states. The Eq. (\ref{eq2}) defines the ``magnetization per spin'' of the system. In addition, we also considered other quantities of interest, namely the susceptibility $\chi$ and the Binder cumulant $U$, defined respectively as
\begin{eqnarray} \label{eq3}
\chi & = &  L^{D}\,(\langle m^{2}\rangle - \langle m \rangle^{2}) \, ~ \\ \label{eq4}
U &  = &  1 - \frac{\langle m^{4}\rangle}{3\,\langle m^{2}\rangle^{2}} \,.
\end{eqnarray}

We will show in the next section that the independence behavior works as a noise that induces a phase transition in the KEOM with the absence of negative interactions.


\section{Results}

\subsection{2D Square lattice}

Let us start considering the model on square lattices. First one can study the symmetric case $g=1/3$. In this case, all probabilities related to the independent behavior, namely $g$ and $(1-g)/2$ are equal to $1/3$. Thus, the probability that an agent $i$ chooses a given opinion $+1$, $-1$ or $0$ independently of the opinions of his nearest neighbors is $q/3$. For the analysis of the model, we have considered the quantities defined on Eqs. (\ref{eq2})-(\ref{eq4}), with $D=2$.

\begin{figure}[t]
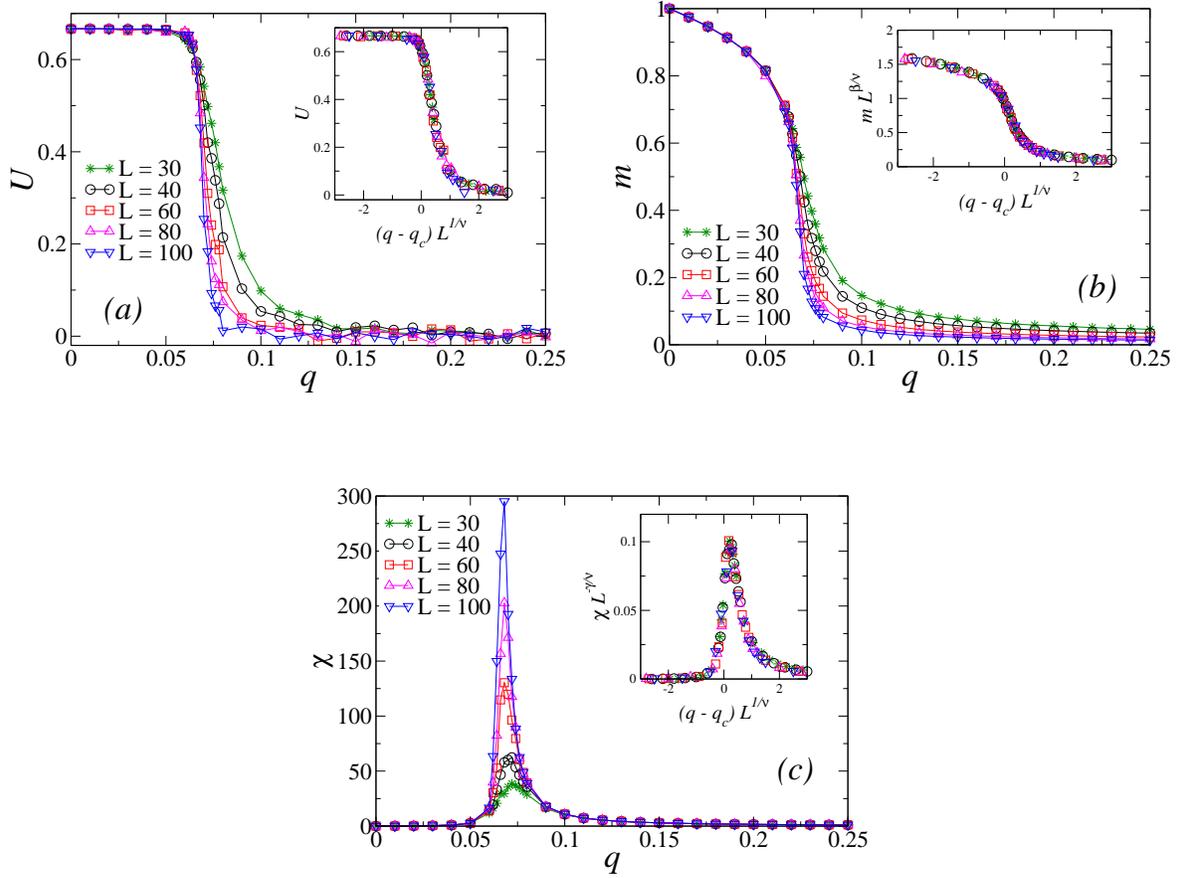

\begin{center}
\vspace{3mm}
\includegraphics[width=0.45\textwidth,angle=0]{figure1a.eps}
\hspace{0.4cm}
\includegraphics[width=0.45\textwidth,angle=0]{figure1b.eps}
\\
\vspace{1.2cm}
\includegraphics[width=0.45\textwidth,angle=0]{figure1c.eps}
\end{center}
\caption{(Color online) Binder cumulant $U$ (a), order parameter $m$ (b) and susceptibility $\chi$ (c) as functions of the independence probability $q$ for the symmetric case ($g=1/3$) on 2D square lattices for different lattice sizes $L$. In the inset we exhibit the corresponding scaling plots. The estimated critical quantities are $q_{c}\approx 0.065$, $\beta\approx 0.125$, $\gamma\approx 1.75$ and $\nu\approx 1.0$. Results are averaged over $300$, $200$, $150$, $120$ and $100$ samples for $L=30, 40, 60, 80$ and $100$, respectively.}
\label{fig1}
\end{figure}

The initial configuration of the population is fully disordered, i.e., we started all simulations with an equal fraction of each opinion ($1/3$ for each one). A time step in the simulations is defined by the application of the rules defined in the previous section $L^{2}$ times. In Fig. \ref{fig1} we exhibit the quantities of interest as functions of $q$ for different lattice sizes $L$. All results suggest the typical behavior of a phase transition. In order to estimate the transition point $q_{c}$, we look for the crossing of the Binder cumulant curves for the different sizes \cite{binder}. From Fig. \ref{fig1} (a), the estimated value is $q_{c}=0.065 \pm 0.003$, where the error bar was determined looking at the crossing of the Binder cumulant curves near $q_{c}$. In addition, in order to determine the critical exponents associated with the phase transition we performed a finite-size scaling (FSS) analysis. We have considered the standard FSS equations, 
\begin{eqnarray} \label{eq5}
m(L) & \sim & L^{-\beta/\nu} \\  \label{eq6}
\chi(L) & \sim & L^{\gamma/\nu} \\   \label{eq7}
U(L) & \sim & {\rm constant} \\   \label{eq8}
q_{c}(L) - q_{c} & \sim & L^{-1/\nu} ~,
\end{eqnarray}

\begin{figure}[t]
\begin{center}
\vspace{0.5cm}
\includegraphics[width=0.45\textwidth,angle=0]{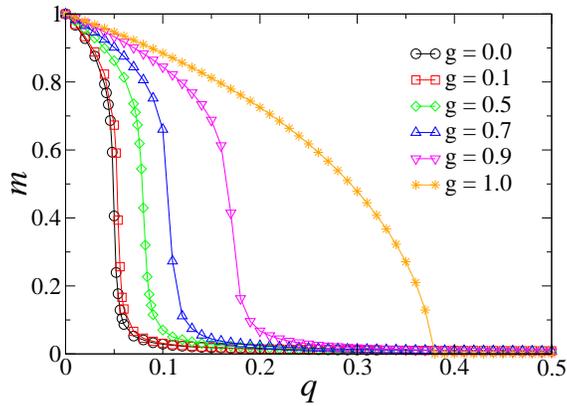}
\end{center}
\caption{(Color online) Order parameter $m$ as a function of $q$ for the model on 2D square lattices for $L=100$ and typical values of $g$. One can see that the transition points depend on $g$. Results are averaged over $100$ simulations.}
\label{fig2}
\end{figure}

\noindent
that are valid in the vicinity of the transition. Thus, we exhibit in the insets of Fig. \ref{fig1} the scaling plots of the quantities of interest ($U$, $m$ and $\chi$). Our estimates for the critical exponents are $\beta=0.125\pm 0.003$, $\gamma=1.75\pm 0.02$ and $\nu=1.00\pm 0.05$, where the error bars were determined by the monitoring small fluctuations around the best data collapse. These values are compatible with the exponents of the two-dimensional Ising model \cite{gould}, suggesting the same univesality class, as equally observed in the mean-field case \cite{meu_indep}. However, there is an important difference. In the mean-field case it was observed that the critical probability ($q_{c}=1/4$) presents the same value of the critical fraction of negative interactions ($p_{c}=1/4$) of the standard KEOM of Ref. \cite{biswas}. In other words, the inclusion of independence with symmetric probabilities (i.e., $g=1/3$) in the fully-connected case produces the same effect of the introduction of negative interactions. In our case, considering negative interactions ($p>0$) with no independence in the square lattice, the model undergoes the order-disorder transition at $p_{c}\approx 0.134$ \cite{mukherjee}. Nevertheless, the critical independence probability in the absence of competitive interactions for the symmetric case was found to be $q_{c}\approx 0.065$, different of $p_{c}$. This difference can be viewed as effects of correlations due to the presence of neighbors, that do not exist in the mean-field case where each agent can interact with all others.

One can also consider the general case where $g\neq 1/3$. In this case, for an agent that act independently, the probabilities to choose the three possible opinions are different. As was done before, we started all simulations with an equal fraction of each opinion. In Fig. \ref{fig2} we show the order parameter as a function of $q$ for typical values of the flexibility $g$ and lattice size $L=100$. One can see that the phase transition occurs for all values of $g$ exhibited in Fig. \ref{fig2}, and the critical points depend on $g$, i.e., we have $q_{c}=q_{c}(g)$. Furthermore, another interesting result that one can see in Fig. \ref{fig2} is that for $g=1$ the order parameter goes exactly to $m=0$, presenting no finite-size effects as the other curves for $g<1$. This result also occurs in the mean-field approximation \cite{meu_indep}, and can be easily understood. Indeed, for $g=1$ all agents that behave independently choose opinion $o=0$. Thus, for a sufficiently large value of $q$ all agents will change independently to $o=0$, which imply we will have $m=0$. This qualitative discussion was confirmed analyticaly in the mean-field case, and will be analyzed numerically in this work.

\begin{figure}[t]
\begin{center}
\vspace{1.0cm}
\includegraphics[width=0.45\textwidth,angle=0]{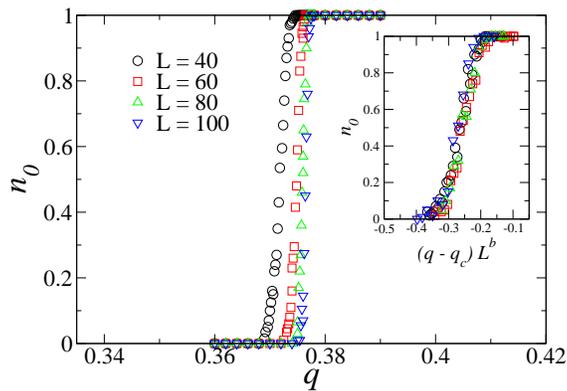}
\end{center}
\caption{(Color online) Fraction $n_{0}$ of samples (over $200$ simulations) that reach the absorbing state with all neutral opinions as a function of $q$ for the model defined on 2D square lattices. Results are for $g=1$ and typical lattice sizes $L$ (main plot). In the inset it is exhibited the corresponding scaling plot. The best collapse of data was obtained for $q_{c}=0.38$ and $b=0.95$.}
\label{fig3}
\end{figure}

\begin{figure}[t]
\begin{center}
\vspace{1.0cm}
\includegraphics[width=0.5\textwidth,angle=-90]{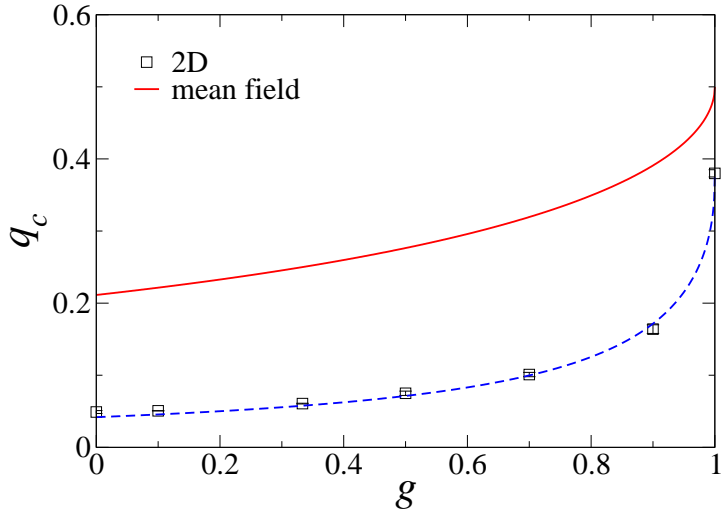}
\end{center}
\caption{(Color online) Comparative phase diagram of the model in the plane $q_{c}$ versus $g$, separating the ordered and the disordered phases. The symbols are the numerical estimates of the critical points $q_{c}$ for the 2D case, whereas the dashed line is a sketch of the boundary, given by Eq. (\ref{eq10}). It is also shown the mean-field result \cite{meu_indep} (full line). The error bars determined by the FSS analysis are smaller than data points.}
\label{fig4}
\end{figure}

Thus, the case $g=1$ is special, because all agents change their opinions to $o=0$ for a sufficent large value of $q$. Indeed, if all agents are in the $o=0$ state, the evolution equation (\ref{eq1}), when applied (with probability $1-q$), does not change the opinions to $+1$ or $-1$ anymore, which means that the system is in an absorbing state. This fact, together with the absence of finite-size effects for the magnetization per spin defined in Eq. (\ref{eq2}), suggests that one can not apply the FSS equations (\ref{eq5}) - (\ref{eq8}). In this case, it is better to analyze other quantity as an order parameter, as was done for the mean-field case \cite{meu_indep}. Thus, following \cite{meu_indep}, we performed several simulations of the system for $g=1$ and we measured the fraction $n_{0}$ of samples that reached the absorbing state with all opinions $0$ as a function of $q$. The result is exhibited in Fig. \ref{fig3} for typical values of $L$, and in this case this order parameter depends on the system size. Considering scaling equations in a similar way as in \cite{meu_indep}, i.e., plotting $n_{0}$ as a function of the variable $(q - q_{c})\,L^{b}$, one obtains $q_{c}=0.38 \pm 0.001$, in agreement with the previous discussion, and $b=0.95 \pm 0.02$. The corresponding data collapse is exhibited in the inset of Fig. \ref{fig3}. Thus, for a sufficient large system, considering $g=1$, for $q>0.38$ all agents will be in the neutral state.

As above discussed, the numerical results suggest that critical points $q_{c}$ depend on $g$. We performed a FSS analysis based on Eqs. (\ref{eq5}) - (\ref{eq8}) in order to obtain the critical points and the critical exponents for other values of $g<1$. In Fig. \ref{fig4} the numerical estimates of $q_{c}(g)$ are plotted as well as the comparison with the mean-field result (see the full line in Fig. \ref{fig4}), given by \cite{meu_indep}

\begin{equation} \label{eq9}
q_{c}(g) = q_{c}(1)\,\left[1 - \left(\frac{1-g}{3-g}\right)^{1/2} \right] \, .
\end{equation}
\noindent
where $q_{c}(1)=1/2$ for the mean-field case. Based on the above equation, we propose the following boundary for the 2D case,
\begin{equation} \label{eq10}
q_{c}(g) = q_{c}(1)\,\left[1 - \left(\frac{1-g}{c_{1}-g}\right)^{c_{2}} \right] \, ,
\end{equation}
\noindent
with two fitting parameters $c1$ and $c2$, and we have used our numerical estimate $q_{c}(1)= 0.38$. Fitting the data, we obtained $c_{1}\approx 1.325$ and $c_{2}\approx 0.415$. Eq. (\ref{eq10}) with these fitted parameters is plotted in Fig. \ref{fig4} (see the dashed line), and one can see a good agreement with the numerical data. In addition, the critical exponents are the same for all values of $g<1$, i.e., we have $\beta\approx 0.125$, $\gamma\approx 1.75$ and $\nu\approx 1.0$, which indicates a universality on the order-disorder frontier of the model, except on the ``special'' point $g=1$ where we have an active-absorbing transition.


\subsection{3D Cubic lattice}

In order to discuss about the robbustness of the universality class of the model, we also considered simulations on cubic lattices with $N=L^{3}$ sites. For simplicity, we only considered the symmetric case with $g=1/3$. However, as one saw in the last subsection, the exponents did not changed for distinct values of $g$ (except for the case $g=1$).

\begin{figure}[t]
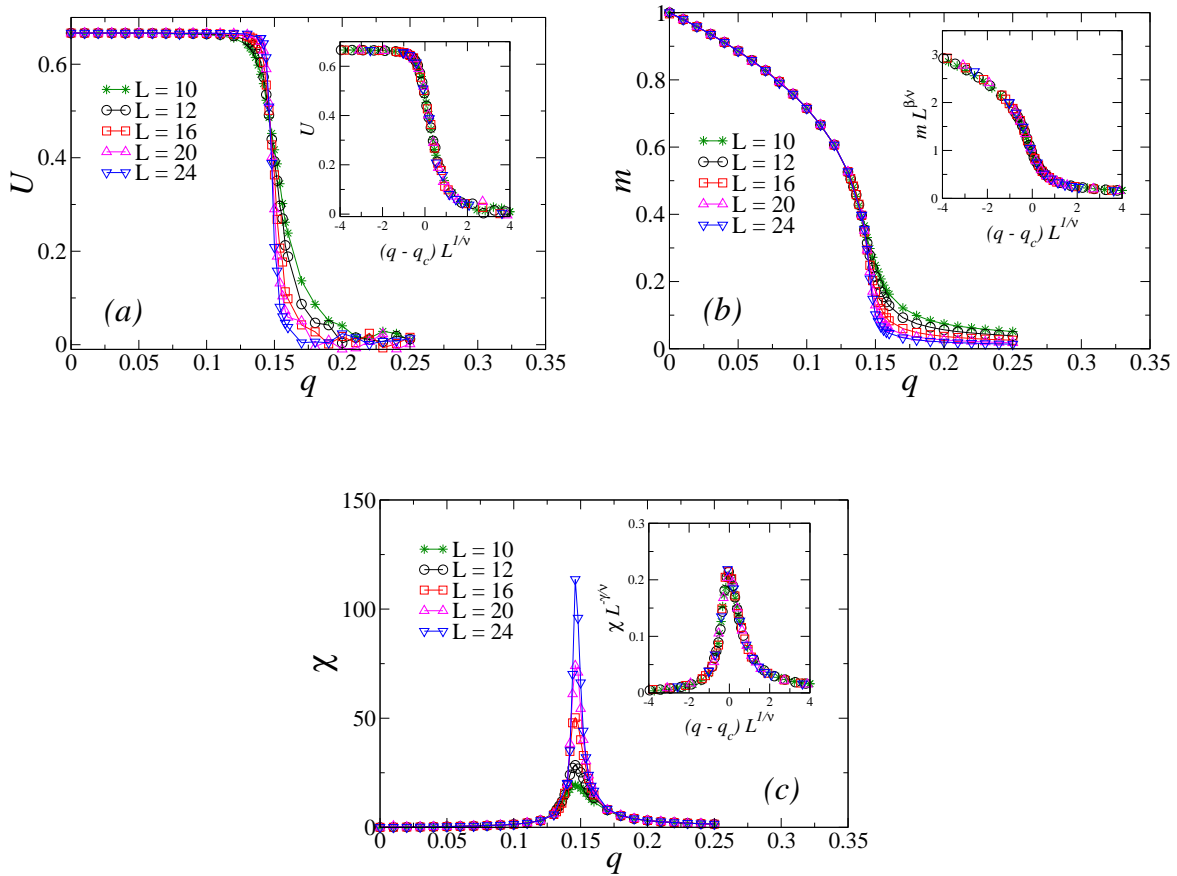

\begin{center}
\vspace{3mm}
\includegraphics[width=0.45\textwidth,angle=0]{figure5a.eps}
\hspace{0.4cm}
\includegraphics[width=0.45\textwidth,angle=0]{figure5b.eps}
\\
\vspace{1.2cm}
\includegraphics[width=0.45\textwidth,angle=0]{figure5c.eps}
\end{center}
\caption{(Color online) Binder cumulant $U$ (a), order parameter $m$ (b) and susceptibility $\chi$ (c) as functions of the independence probability $q$ for the symmetric case ($g=1/3$) on 3D cubic lattices for different lattice sizes $L$. In the inset we exhibit the corresponding scaling plots. The estimated critical quantities are $q_{c}\approx 0.146$, $\beta\approx 0.32$, $\gamma\approx 1.23$ and $\nu\approx 0.62$. Results are averaged over $300$, $200$, $150$, $120$ and $100$ samples for $L=10, 12, 16, 20$ and $24$, respectively.}
\label{fig5}
\end{figure}

Again, we considered as the initial configuration a fully-disordered population. A time step in the simulations is defined by the application of the rules defined in the previous section $L^{3}$ times. In Fig. \ref{fig5} we exhibit the quantities of interest as functions of $q$ for different lattice sizes $L$. All results suggest the typical behavior of a phase transition. Again, we can look for the crossing of the Binder cumulant curves for the different sizes \cite{binder} to estimate the critical point $q_{c}$. From Fig. \ref{fig5} (a), the estimated value is $q_{c}=0.146 \pm 0.002$. In addition, in order to determine the critical exponents associated with the phase transition we performed a FSS analysis. We have considered the standard FSS equations (\ref{eq5}) - (\ref{eq8}). Thus, we exhibit in the insets of Fig. \ref{fig5} the scaling plots of the quantities of interest ($U$, $m$ and $\chi$). Our estimates for the critical exponents are $\beta=0.32\pm 0.01$, $\gamma=1.23\pm 0.02$ and $\nu=0.62\pm 0.02$, that are compatible with the exponents of the three-dimensional Ising model \cite{gould}, confirming the Ising model univesality class. However, as in the 2D case, the critical noise ($q_{c}\approx 0.146$) is different from the critical fraction of negative interactions observed in the square lattice, $p_{c}\approx 0.199$ \cite{biswas}.


\begin{figure}[t]
\begin{center}
\vspace{3mm}
\includegraphics[width=0.45\textwidth,angle=0]{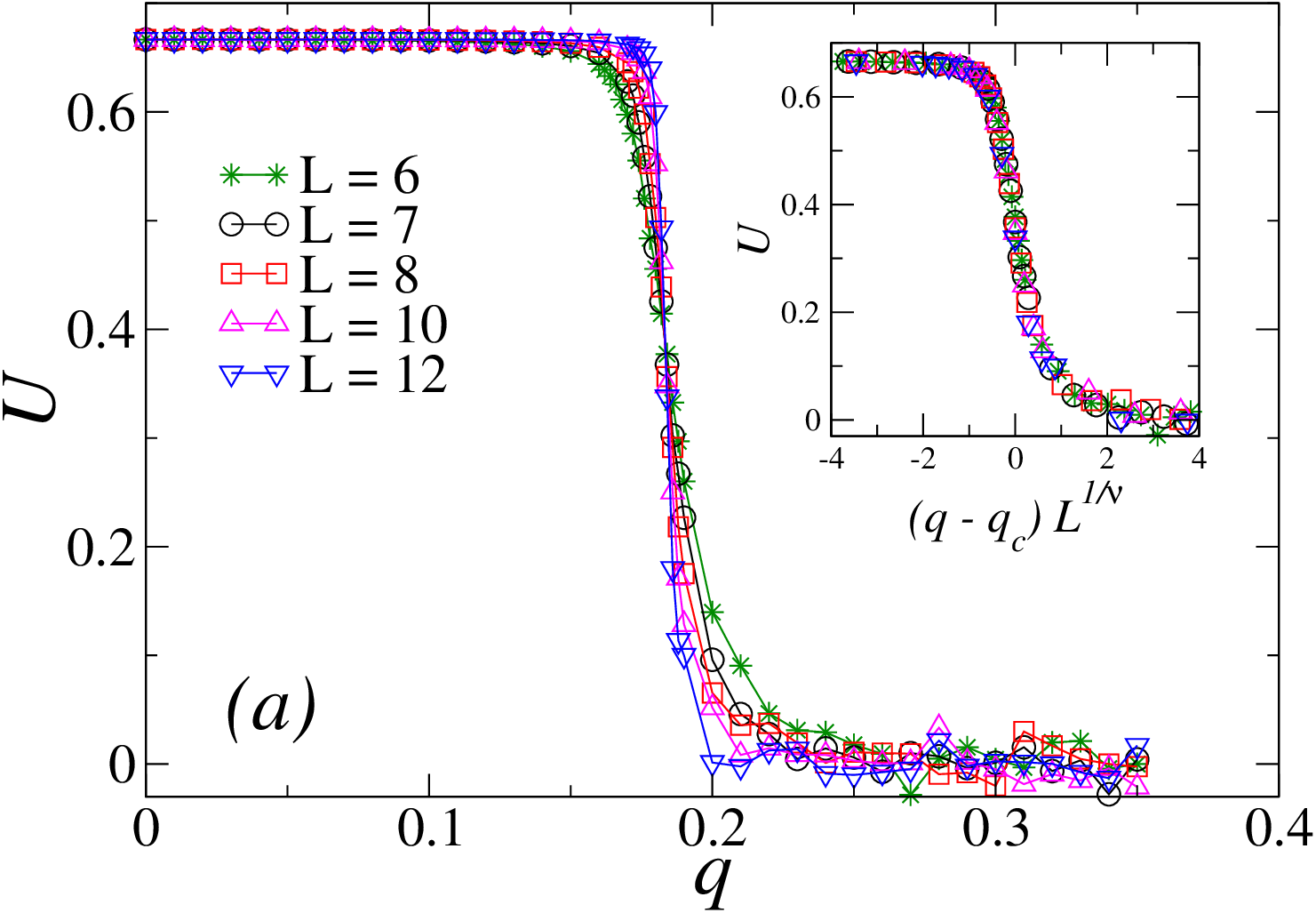}
\hspace{0.4cm}
\includegraphics[width=0.45\textwidth,angle=0]{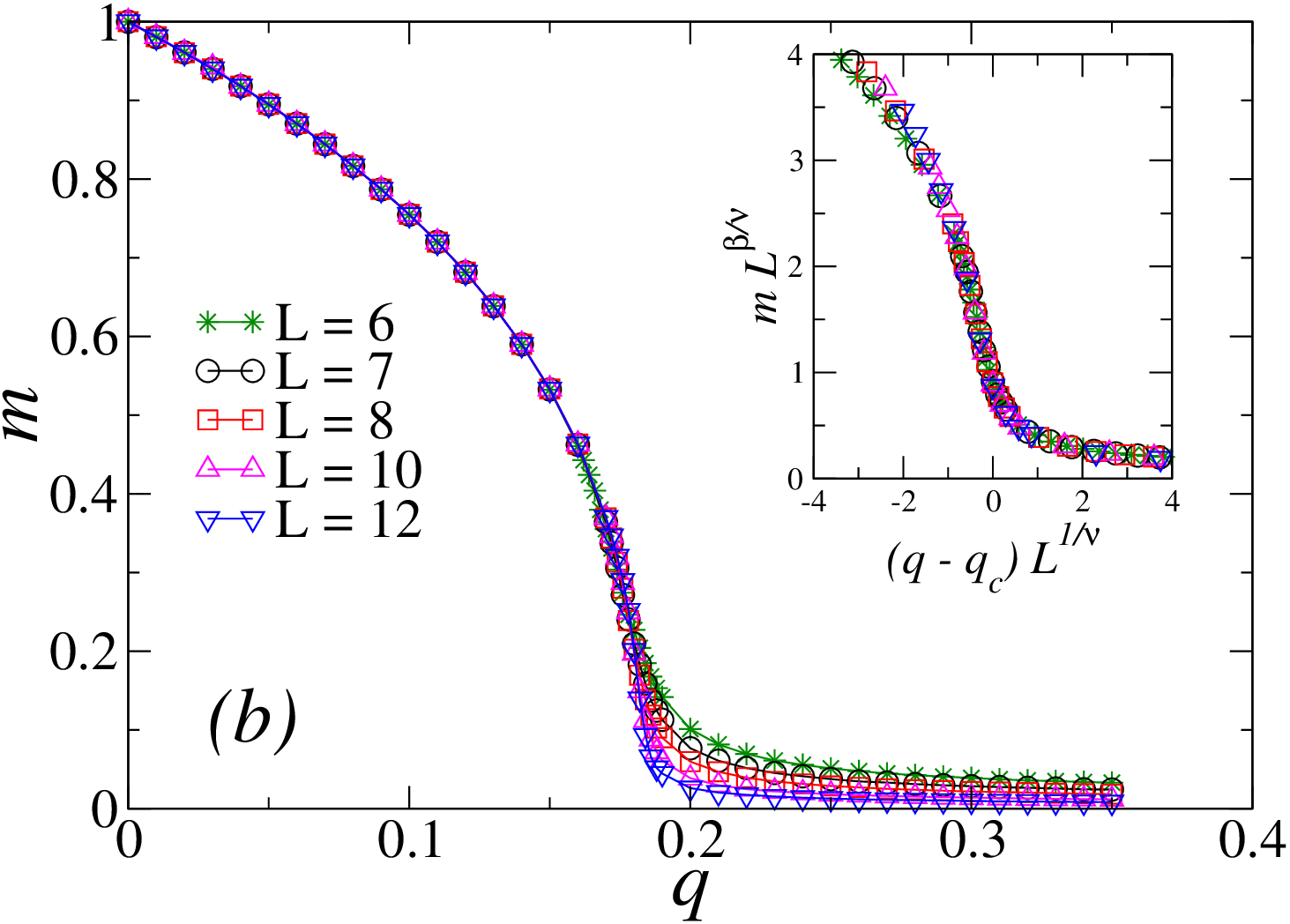}
\\
\vspace{1.2cm}
\includegraphics[width=0.45\textwidth,angle=0]{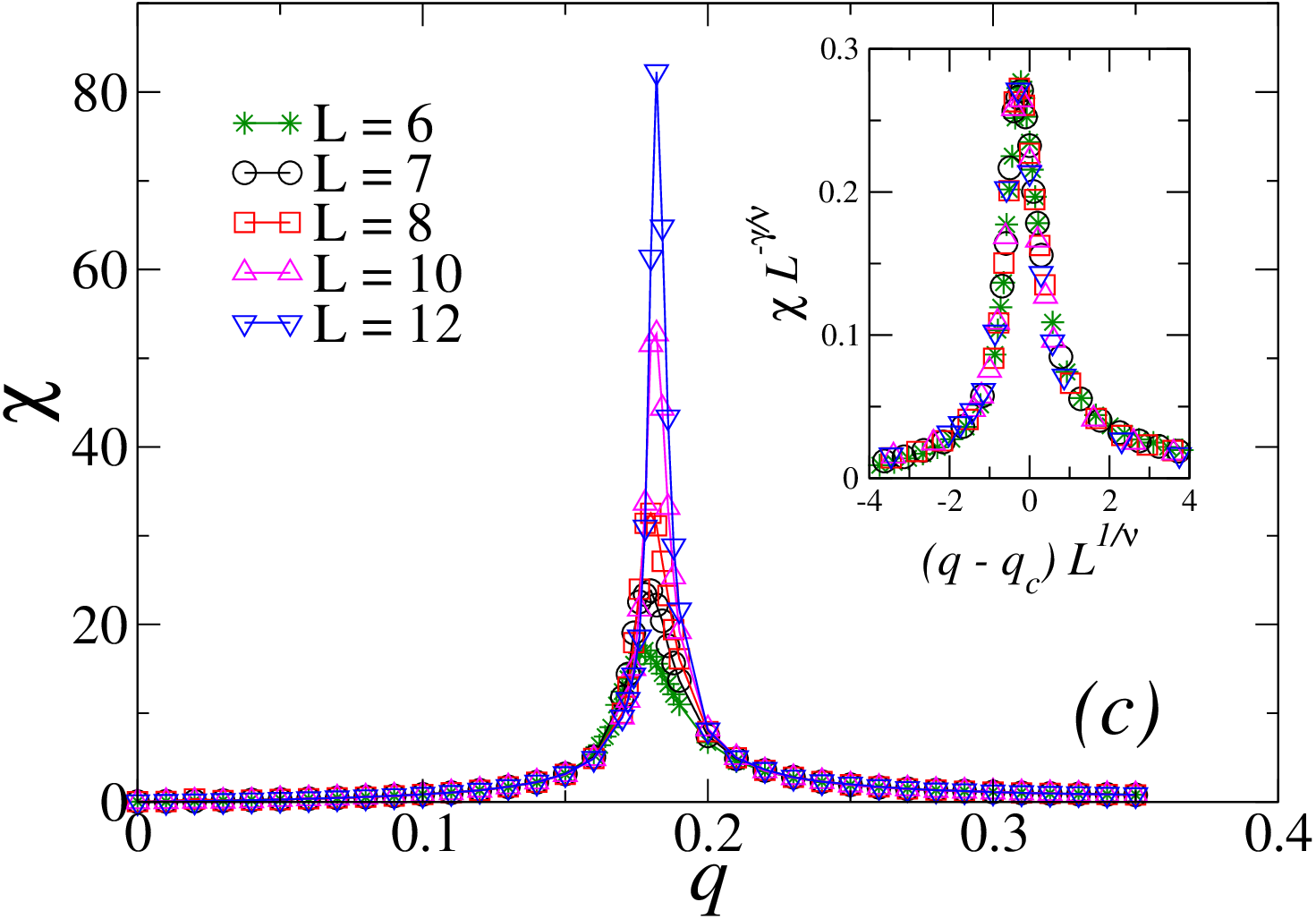}
\end{center}
\caption{(Color online) Binder cumulant $U$ (a), order parameter $m$ (b) and susceptibility $\chi$ (c) as functions of the independence probability $q$ for the symmetric case ($g=1/3$) on 4D hypercubic lattices for different lattice sizes $L$. In the inset we exhibit the corresponding scaling plots. The estimated critical quantities are $q_{c}\approx 0.183$, $\beta\approx 0.48$, $\gamma\approx 1.02$ and $\nu\approx 0.51$. Results are averaged over $200$, $170$, $120$, $100$ and $100$ samples for $L=6, 7, 8, 10$ and $12$, respectively.}
\label{fig6}
\end{figure}

\subsection{4D Hypercubic lattice}

We also considered the model on hypercubic four-dimensional lattices with $N=L^{4}$ sites. As in the previous case (3D), we only considered the symmetric case with $g=1/3$.

Again, we considered as the initial configuration a fully-disordered population. A time step in the simulations is defined by the application of the rules defined in the previous section $L^{4}$ times. In Fig. \ref{fig6} we exhibit the quantities of interest as functions of $q$ for different lattice sizes $L$. All results suggest the typical behavior of a phase transition. Again, we can look for the crossing of the Binder cumulant curves for the different sizes \cite{binder} to estimate the critical point $q_{c}$. From Fig. \ref{fig6} (a), the estimated value is $q_{c}=0.183 \pm 0.002$. In addition, in order to determine the critical exponents associated with the phase transition we performed a FSS analysis. We have considered the standard FSS equations (\ref{eq5}) - (\ref{eq8}). Thus, we exhibit in the insets of Fig. \ref{fig6} the scaling plots of the quantities of interest ($U$, $m$ and $\chi$). Our estimates for the critical exponents are $\beta=0.48\pm 0.04$, $\gamma=1.02\pm 0.03$ and $\nu=0.51\pm 0.03$, that are compatible with the exponents of the four-dimensional Ising model \cite{cardy}, confirming the Ising model univesality class. Furthermore, notice that the obtained exponents are the same of the mean-field exponents (with exception of $\nu$, as discussed in the subsction II), suggesting that the upper critical dimension of the model is $D_{c}=4$, as for the Ising model \cite{cardy}. A summary of all critical values is exhibited in Table \ref{Tab2}.

\begin{table*}[tbp]
\begin{center}
\renewcommand\arraystretch{1.3} 
\begin{tabular}{|c|c|c|c|c|}
\hline
$D$ & $q_{c}$ & $\beta$ & $\gamma$ & $\nu$     \\ \hline
2 & 0.065\,$\pm$\,0.002 & 0.125\,$\pm$\,0.003 & 1.75\,$\pm$\,0.02 & 1.00\,$\pm$\,0.05  \\ 
3 & 0.146\,$\pm$\,0.003  & 0.32\,$\pm$\,0.01 & 1.23\,$\pm$\,0.02 & 0.62\,$\pm$\,0.02 \\ 
4 & 0.183\,$\pm$\,0.002 & 0.48\,$\pm$\,0.04 & 1.02\,$\pm$\,0.03 & 0.51\,$\pm$\,0.03  \\
mean field & 0.25 & 0.5 & 1.0 & 2.0 \\ \hline
\end{tabular}%
\end{center}
\caption{Critical points $q_{c}$ and the critical exponents $\beta$, $\gamma$ and $\nu$ for distinct lattice dimensions. The mean-field values were obtained form Ref. \cite{biswas}.}
\label{Tab2}
\end{table*}


\section{Comments}   

In this work we introduce the mechanism of independence in a three-state ($+1$, $-1$ and $0$) kinetic exchange opinion model defined on regular D-dimensional lattices. In the absence of negative interactions, this model always evolve to ordered consensus states. Our results show that independence acts as a noise or social temperature, inducing a nonequilibrium phase transition in the model.

For the 2D case, we verified numerically that the critical points depend on the agents' flexibility $g$. The numerical simulations suggest that we have the same critical exponents for all values of $g<1$, i.e., we have $\beta\approx 0.125$, $\gamma\approx 1.75$ and $\nu\approx 1.0$, which suggests a universality on the order-disorder frontier of the model. In addition, the model presents the same universality class of the equilibrium Ising model on a square lattice. On the other hand, the case $g=1$ is special, and the system undergoes a transition to an absorbing state with all opinions equal to $0$.

For 3D and 4D cases we analyzed only the symmetric case $g=1/3$. Our estimates for the critical exponentes are consistent with the values for the 3D and 4D Ising models, respectively. In addition, we found for D=4 the same exponents observed in the mean-field formulation of the model, suggesting also that the upper critical dimension of our model is $D_{c}=4$, as for the Ising model.

From the social point of view, we observed that the smaller the dimension D, the smaller the critical value of the independence $q_{c}$. In other words, in structures of social interactions with a small number of connections it is harder to reach a decision on the debate under discussion, i.e., it is hard to observe a majority of one of the sides ($+1$ or $-1$). Thus, even a small fraction of independent behaviors leads the debate to an indecision (disordered state). However, for increasing number of social connections (increasing D), the competition between social interaction and independence (noise) increases, with an advantage to the social pressure, leading to a larger value of $q$ needed to disorder the system. This fact implies that it is easier to observe a majority (debate with a decision, or an ordered state).

Due to the previous discussion, it is interesting to analyze the model on complex networks, that better represent the modern networks of social interactions. The presence of a topology certainly affects the critical behavior of the system. An evidence for that is given by the present paper, were we observe distinct exponents in 2D and 3D in comparison with the mean-field case. Thus, the consideration of a complex network for the social interactions probably will change the critical exponents in comparison with the estimated in the paper, but it goes beyond the target of this work. It will certainly be considered in a future work.


\section*{Acknowledgments}

The author acknowledges financial support from the Brazilian Scientific Funding Agency CNPq.


\begin{thebibliography}{40}

\bibitem{galam_book}
S. Galam, \textit{Sociophysics: A Physicist's Modeling of Psycho-political Phenomena} (Springer, Berlin, 2012).

\bibitem{sen_book}
P. Sen, B. K. Chakrabarti, \textit{Sociophysics: an introduction} (Oxford University Press, Oxford, 2013).


\bibitem{galam}
S. Galam, Physica A \textbf{333} (2004) 453-460.

\bibitem{lalama}
M. S. de la Lama, J. M. Lopez, H. S. Wio, Europhys. Lett. \textbf{72} (2005) 851-857.

\bibitem{sznajd_indep1}
K. Sznajd-Weron, M. Tabiszewski, A. M. Timpanaro, Europhys. Lett. \textbf{96} (2011) 48002;


\bibitem{sznajd_indep2}
P. Nyczka, K. Sznajd-Weron, J. Cislo, Phys. Rev. E \textbf{86}, 011105 (2012).


\bibitem{sznajd_indep3}
P. Nyczka, K. Sznajd-Weron, J. Stat. Phys. \textbf{151} (2013) 174-202.


\bibitem{javarone1}
S. Galam, M. A. Javarone, PLoS ONE \textbf{11}(5): e0155407 (2016).

\bibitem{javarone2}
M. A. Javarone, T. Squartini, J. Stat. Mech. P10002 (2015).

\bibitem{meu_indep}
N. Crokidakis, Phys. Lett. A \textbf{378} (2014) 1683.

\bibitem{meu_pmco}
N. Crokidakis, P. M. C. de Oliveira, Phys. Rev. E \textbf{92}, 062122 (2015).


\bibitem{lccc}
M. Lallouache, A. S. Chakrabarti, A. Chakraborti, B. K. Chakrabarti, Phys. Rev. E \textbf{82}, 056112 (2010).

\bibitem{p_sen}
P. Sen, Phys. Rev. E \textbf{83} (2011) 016108.

\bibitem{biswas11} S. Biswas, Phys. Rev. E \textbf{84}, 056106 (2011).

\bibitem{biswas}
S. Biswas, A. Chatterjee, P. Sen, Physica A \textbf{391} (2012) 3257-3265.


\bibitem{mukherjee}
S. Mukherjee, A. Chatterjee, Phys. Rev. E \textbf{94}, 062317 (2016).
  

\bibitem{binder}
K. Binder, Z. Phys. B \textbf{43}, 119 (1981).


\bibitem{gould}
H. Gould, J. Tobochnik, \textit{An Introduction to Computer Simulation Methods: Applications to Physical Systems} (Addison-Wesley, Massachusetts, 1996).

\bibitem{cardy}
J. Cardy, \textit{Scaling and Renormalization in Statistical Physics} (Cambridge University Press, Cambridge, 2002).
  



\end{thebibliography}
\end{document}